\documentclass[12pt]{iopart}
\usepackage{graphicx}

\begin{document}

\title[Dependences of the Casimir-Polder interaction
on atomic and material properties]{Dependences of 
the Casimir-Polder interaction
between an atom and a cavity wall on atomic and
material properties
}
\author{V~M~Mostepanenko$^{1}$,
J~F~Babb$^{2}$,
 A~O~Caride$^{3}$, 
G~L~Klimchitskaya$^{4}$ and S~I~Zanette$^{3}$
}

\address{$^1$
Noncommercial Partnership  ``Scientific Instruments'', Moscow, Russia}
\address{$^2$
Institute for Theoretical Atomic, Molecular and Optical Physics,
Harvard-Smithsonian Center for Astrophysics, Cambridge,
MA 02138, USA}
\address{$^3$
Centro Brasileiro de Pesquisas F\'{\i}sicas, Rio de Janeiro,
RJ 22290-180, Brazil}
\address{$^4$
North-West Technical University, Millionnaya St. 5,
St.Petersburg, Russia
}

\begin{abstract}
The Casimir-Polder and van der Waals interactions
 between an atom and a flat
cavity wall are investigated under the influence of real
conditions including the dynamic polarizability of the
atom, actual conductivity of the wall material and
nonzero temperature of the wall. The cases of different
atoms near metal
and dielectric walls are considered. It is shown that
to obtain accurate results for the atom-wall
interaction at short separations,
one should use the complete tabulated optical data for
the complex refractive index of the wall material and the
accurate dynamic polarizability of an atom. At relatively
large separations in the case of a metal wall, one may use
the plasma model dielectric function to describe the
dielectric properties of wall material.  The obtained
results are important for the theoretical interpretation
of experiments on quantum reflection and Bose-Einstein
condensation.
\end{abstract}
%\submitto{\JPA}
%\maketitle

\section*{}
Recently the study of dispersion interactions between
an atom and a wall has assumed a new significance in connection
with Bose-Einstein condensates of ultracold atoms [1--3].
The van der Waals and Casimir-Polder forces 
acting between dilute individual
atoms, confined in a magnetic trap, and a wall may influence the
stability of a condensate and the effective size of the trap
\cite{3}. As was shown in Ref.~\cite{4}, the study of the
collective oscillations of the Bose-Einstein condensate can
provide a sensitive test of dispersion forces. This
prediction was later supported both theoretically \cite{5} and
experimentally \cite{6}. Dispersion interaction between
an atom and a wall is also taken into account in quantum
reflection of cold atoms on a surface \cite{7} and in dynamical
interaction effects of fast atoms and molecules with solid
surfaces \cite{8}. Currently the new asymptotic behavior of
the surface-atom interaction out of thermal equilibrium has 
been advanced \cite{9}. Below we use the generic name
``Casimir-Polder" for all atom-wall interactions of dispersion
nature because the pure nonretarded regime occurs at separations
from zero to a few nanometers only.

The theoretical basis for the description of the Casimir-Polder
interaction between an atom at a separation $a$ from a flat
wall at temperature $T$ in thermal equilibrium is given by the
Lifshitz-type formula for the free energy [10--12]
\begin{eqnarray}
&&
{\cal F}(a,T)=-\frac{k_B T}{8a^3}\left\{
\vphantom{\int_{\zeta_l}^{\infty}}
2\alpha(0)f(0)+\sum\limits_{l=1}^{\infty}
\alpha(i\zeta_l\omega_c)\right.
\label{eq1} \\
&&
\phantom{aaaaaa}\left.\times
\int_{\zeta_l}^{\infty}dye^{-y}\left[
\left(2y^2-\zeta_l^2\right)r_{\|}(\zeta_l,y)+
\zeta_l^2r_{\bot}(\zeta_l,y)\right]
\vphantom{\sum\limits_{l=1}^{\infty}}\right\}.
\nonumber
\end{eqnarray}
\noindent
Here $\alpha(\omega)$ is the atomic dynamic polarizability,
$k_B$ is the Boltzmann constant, $\zeta_l=4\pi lk_BTa/(\hbar c)$
are the dimensionless Matsubara
frequencies, $\omega_c=c/(2a)$ is the characteristic
frequency of the Casimir-Polder interaction, and the reflection
coefficients for two independent polarizations of electromagnetic
field are defined as
\begin{equation}
r_{\|}(\zeta_l,y)=\frac{\varepsilon_ly-
\sqrt{y^2+\zeta_l^2(\varepsilon_l-1)}}{\varepsilon_ly+
\sqrt{y^2+\zeta_l^2(\varepsilon_l-1)}},
\quad
r_{\bot}(\zeta_l,y)=\frac{\sqrt{y^2+\zeta_l^2(\varepsilon_l-1)}
-y}{\sqrt{y^2+\zeta_l^2(\varepsilon_l-1)}+y},
\label{eq1a}
\end{equation}
\noindent
where $\varepsilon_l\equiv\varepsilon(i\zeta_l\omega_c)$ is
the permittivity of wall material computed at
imaginary Matsubara frequencies. For dielectrics
$f(0)=[\varepsilon(0)-1]/[\varepsilon(0)+1]$ and for metals
$f(0)=1$.

In most  calculations of the atom-wall interaction previously performed
only the limiting cases of large and short
separations were considered. The polarizability of the atom
was taken into account in the static approximation \cite{13} or
in the framework of the single-oscillator model \cite{14},
and the dielectric properties of the wall material were
oversimplified (for example, by considering a metal wall to be 
made of ideal metal). The present experimental situation 
requires precise (1\% accuracy) computations of the 
Casimir-Polder interaction in a wide separation range
from about 3\,nm (where the Lifshitz formula becomes
applicable) to 10\,$\mu$m. In this paper we present the
results of such computations clarifying the atomic and material
properties which are essential to attain the required
accuracy. 

We have performed numerical computations of the free-energy
(\ref{eq1}), (\ref{eq1a}) for metastable He${}^{\ast}$, Na, and Cs atoms
in ground state located near metal (Au), semiconductor (Si) and
dielectric (SiO${}_2$) walls at $T=300\,$K. 
(The modification on account of walls in the spontaneous emission
of Rydberg atoms, obtained, e.g., by means of two lasers, is discussed
in Refs.~\cite{14a,14b}. However, thermal quanta at $T=300\,$K are too
small to excite atom from the ground state to some other states.)
Three different 
descriptions for the dielectric properties of a metal were
used: \textit{i}) as an ideal metal, 
\textit{ii}) using the dielectric permittivity from the 
free-electron plasma model 
$\varepsilon(i\xi)=1+\omega_p^2/\xi^2$ (where $\omega_p$
is the plasma frequency), and \textit{iii})
with $\varepsilon(i\xi)$ obtained
by means of dispersion relation using the tabulated
optical data for the complex index of refraction \cite{15}.
The dielectric permittivity of a semiconductor or
dielectric was described either by their static permittivity
 $\varepsilon(0)$ or by means of their tabulated optical data 
and the dispersion relation. The polarizability of an atom was 
represented by its static value $\alpha(0)$ or by means of the 
highly accurate $N$-oscillator model \cite{16}
\begin{equation}
\alpha(i\zeta_l\omega_c)=\frac{e^2}{m}
\sum\limits_{n=1}^{N}
\frac{f_{0n}}{\omega_{0n}^2+\omega_c^2\zeta_l^2},
\label{eq2}
\end{equation}
\noindent
where $m$ and $e$ are the electron mass and charge,
$f_{0n}$ and $\omega_{0n}$ are the oscillator strength 
and frequency of the $n$th excited-state to ground-state 
transition, respectively. A more simplified
single-oscillator model [Eq.~(\ref{eq2}) with $N=1$] was also used.

%%%%%%%%%%%%%%%%%%%%%%%%%%%%%%%%%%%%%%%%%%%%%
\begin{table}
\caption{\label{label}Free energy ${\cal F}$ (in J) of the Casimir-Polder
interaction between a He${}^{\ast}$ atom and Au and SiO${}_2$ walls
[columns (a)] and correction factors to it at different
separations $a$. In columns labeled (a) the 
material of the wall and the atom
are described by the optical tabulated data and accurate dynamic
polarizability, respectively. In columns labeled (b)
the metal is an ideal
one and the dielectric permittivity of SiO${}_2$ is static; 
the dynamic polarizability of the atom 
is the accurate one. In columns 
labeled (c) the wall materials are described by the 
tabulated optical data and the dynamic polarizability 
of the atom is given by the
single-oscillator model. In column (d)  the metal is described by the
plasma model and the dynamic polarizability of
the atom is accurate.}
\begin{indented}
\item[]\begin{tabular}{@{}rccccccc}
\br
&\centre{4}{He$^{\ast}$ near a Au wall}
&\centre{3}{He$^{\ast}$ near a SiO${}_2$ wall} \\
\ns&\crule{4}&\crule{3}\\
{$\phantom{aa}$ }$a\,$(nm){$\phantom{a}$ }&(a)&(b)&(c)&(d)
&(a)&(b)&(c) \\
\mr
3{$\phantom{aaa}$ } & 3.80$\times 10^{-23}$ & 1.16 & 0.956 & 0.937 &
1.61$\times 10^{-23}$ & 1.78 & 0.949 \\
10{$\phantom{aaa}$} &9.95$\times 10^{-25}$ &1.14&0.961&0.948&
4.18$\times 10^{-25}$ & 1.73 & 0.958 \\
20{$\phantom{aaa}$} &1.18$\times 10^{-25}$ &1.14&0.973&0.959&
4.94$\times 10^{-26}$ & 1.68 & 0.967 \\
50{$\phantom{aaa}$} &6.62$\times 10^{-27}$ &1.13&0.984&0.976&
2.71$\times 10^{-27}$ & 1.64 & 0.983 \\
100{$\phantom{aaa}$} &6.98$\times 10^{-28}$ &1.11&0.991&0.981&
2.76$\times 10^{-28}$ & 1.60 & 0.993 \\
150{$\phantom{aaa}$} &1.77$\times 10^{-28}$ &1.10&0.997&0.992&
6.93$\times 10^{-29}$ & 1.57 & 0.994 \\
\br
\end{tabular}
\end{indented}
\end{table}
%%%%%%%%%%%%%%%%%%%%%%%%%%%%%%%%%%%%%%%%%%%%%%%%%%%%%%%%%%%%%
Computations show that at short separations (from 3\,nm to
about 150\,nm) it is necessary to use the complete tabulated
optical data for the complex index of refraction in order to 
find the most accurate results. For the dynamic
polarizability of an atom, at shortest separations the highly
accurate data for it should be used. With increasing 
atom-wall 
distance up to several tens of nanometers the
single-oscillator model becomes applicable. These calculations
are illustrated in Table 1 by the example of a metastable
He${}^{\ast}$ atom near Au and SiO$_2$ walls (the analogous 
results for Na and Cs atoms near Au, Si, and SiO$_2$ walls
can be found in Refs.~\cite{11,12}). The tabulated optical
data for Au and SiO${}_2$ were taken from Ref.~\cite{16a},
and the values of Au plasma frequency and SiO${}_2$ static
permittivity are 
$\omega_p=9.0\,\mbox{eV}=1.37\times 10^{16}\,$rad/s and 
$\varepsilon(0)=3.84$. The accurate data for the dynamic 
polarizability of metastable He${}^{\ast}$ (with a relative
error of order $10^{-6}$) were taken from Ref.~\cite{16b}
and the parameters of a single-oscillator model
from Ref.~\cite{16c} were used. 
As is seen in Table 1,
the use 
of the ideal metal or the static
dielectric permittivity  approximations leads to errors up to 16\% for metal
and 78\% for dielectric. These errors slowly decrease
with increasing separation between the atom and
the wall. The plasma model is a better
approximation than the ideal metal approximation. It results in errors
of about 5\% at the shortest separations and becomes sufficiently
exact when the separation approaches 150\,nm. The use of the static
atomic polarizability would result in much greater errors 
and for this reason it is omitted from Table 1.

%%%%%%%%%%%%%%%%%%%%%%%%%%%%%%%%%%%%%%%%%%%%%%%%%%%%%%%%%%%%%%
\begin{figure*}[t]
\vspace*{-10cm}
\includegraphics{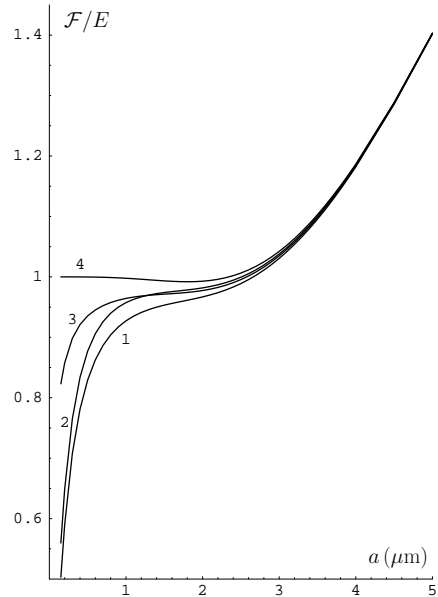}
\vspace*{-11.3cm}
\caption{Normalized Casimir-Polder free energy for 
metastable He${}^{\ast}$
atom near Au wall versus separation. Lines 1,\,2 take into
account the dynamic polarizability of an atom (in the 
single-oscillator model) and describe metal by the plasma
model or as ideal one, respectively.
Lines 3,\,4 describe atom by the static polarizability and metal 
in analogy with lines 1,\,2.
}
\end{figure*}
At large separations, from 150\,nm to a few micrometers, the
effects of the atomic dynamic polarizability play a more important
role than the effects of the finite conductivity of the metal.
The single-oscillator model, however, is sufficient to achieve
the required accuracy. The dielectric properties of a metal
can be approximated by the plasma model.  For  dielectrics
and semiconductors both tabulated optical data and the
Ninham-Parsegian representation for the dielectric permittivity
\cite{17} are suitable for obtaining accurate results. 
For sufficiently large separations one
can use the static dielectric permittivity of the wall. We
illustrate these features using the example of  a  He${}^{\ast}$
atom near an Au wall. Due to the strongly nonmonotonous
dependence of the free energy on separation, we plot along
the vertical axis the ratio of the free energy to the
Casimir-Polder energy $E(a)=-3\hbar c\alpha(0)/(8\pi a^4)$
of an atom near a wall made of ideal metal at $T=0$.
As is seen from Fig.~1, at separations $a>(4-5)\,\mu$m
all approaches lead to approximately equal values
of the free energy.

To conclude,  results such as  those presented in the columns
labeled 
(a) in Table~1 and by  line~1 in Fig.~1 can be used in
interpretation of precision experiments on  atom-surface
interactions.

\section*{Acknowledgments}
ITAMP is supported in part by a grant from the NSF to the Smithsonian
Institution and Harvard University.
 VMM and GLK were partially
supported by FAPERJ (proceess Nos. E--26/170.132 and 170.409/2004)
and by the Russian Foundation for Basic Research (grant 
No. 05--08--18119a).
%%%%%%%%%%%%%%%%%%%%%%%%%%%%%%%%%%%%%%%%%%%%%%%%%%%%%%%%%%%%%%%%%

%%%%%%%%%%%%%%%%%%%%%%%%%%%%%%%%%%%%%%%%%
\end{document}